\documentclass[12pt]{article}
\usepackage{a4}
\usepackage{epsf}
\begin{document}
\begin{center}
{\bf      CAN BARYOGENESIS SURVIVE IN THE STANDARD MODEL
 DUE TO STRONG HYPERMAGNETIC FIELD? }

                                ~\\
                                ~\\

 {Vladimir Skalozub $^a$ and Vadim Demchik$^a$ } \\
                                  ~\\
                                   ~\\
       $^a$ Dniepropetrovsk State University, 320625 Dniepropetrovsk, Ukraine\\
       e-mail:  Skalozub@ff.dsu.dp.ua; dvi@ff.dsu.dp.ua
\\
                         ~\\
                        ~\\
                       ABSTRACT
\end{center}

    The electroweak phase transition in a constant hypermagnetic field  is
studied in
 the Standard Model. The symmetry behaviour is investigated within the
consistent effective potential of the scalar and magnetic fields at finite
temperature. It includes  the one-loop and ring diagram contributions.
All fundamental fermions and bosons are taken into consideration with their actual
masses.
The only
free parameter is the Higgs boson mass which is chosen to be in the
energy interval 75 GeV $\le m_H \le$  115 GeV.  It is found that for the
field strengths $H \sim 10^{22}- 10^{23}$G the electroweak phase
transition is of first order but a baryogenesis condition is not
 satisfied. For stronger fields it  becomes of second
order. Hence it is concluded that the smooth hypermagnetic field does not 
generate the strong first order phase transition and the baryogenesis does
not survive in the Standard Model. The comparison with  the results of other
approaches is done.

\section{Introduction}

Among nowadays problems of high energy physics there are two
ones which, at first glance, are not connected with each other.  These are
the value  of the Higgs boson mass $m_H$ and the  magnetic field strengths
 $H$ which can be present  in the early universe (see
surveys \cite{Enq}, \cite{She}). They are of
paramount importance for particle physics and cosmology. For instance, a
large scale homogeneous hypercharge magnetic field could essentially influence
the type of the electroweak (EW) phase transition making it strong first
order \cite{Shap1}, \cite{Elm1}. An interest to this problem increased
recently when it has been realized that without external fields a standard
 baryogenesis does not hold in the minimal Standard Model (SM).

In Refs. \cite{Shap1}, \cite{Elm1}, \cite{Shap4}, \cite{Lain} the influence
 of the constant hypermagnetic field on the EW phase transition has been
investigated.
In the former one the EP
was computed in a tree approximation and the result that the presence of
$H_Y$ makes the weak first-order phase transition stronger has been derived.
  In Ref. \cite{Elm1} the temperature dependent part of the EP was calculated
in one-loop order  whereas the external field has also been allowed for in
tree approximation. Therein  it was found that for the field strengths $H_Y  > 0.3
- 0.5
T^2$, where $T$ is temperature at the transition, the standard baryogenesis
survives. As this paper is concerned, we  would like to notice that for
the weak first-order phase transition the fluctuations are essential,
the one-loop approximation to the EP is not trusty and the correlation
 corrections must be included \cite{Car}, \cite{Din}, \cite{Kol}.
Moreover, no investigation of the EP curve in strong fields has been
carried out in Ref. \cite{Elm1}. In fact, it was just assumed that the phase
 transition is of first order and the jump of the order parameter is
 as at the zero external field. Furthermore, the role of fermions  at high
temperature and strong fields has not been investigated. The same has
relevance to
the papers \cite{Shap1}, \cite{Shap4}, \cite{Lain}. But as it will be seen
in what follows, fermions significantly influence  a vacuum dynamics in
the environment.

To make a link between recent studying of symmetry behaviour  in the external
hypermagnetic field and the already obtained results for the case of usual
magnetic field
\cite{Rez}, \cite{SVZ1}, we notice that  in the broken phase  $H_Y$  is connected
with $H$ by the relation $H = H_{Y} \cos \theta$, where $\theta$ is the
Weinberg angle. So, all investigations dealing with symmetry behaviour in the
magnetic fields at high temperature are relevant to the considered  case of
 $H_Y$ in the respect of the form of the EP curve at different $T, H_Y$. The
 hypercharge field
influences the scalar field condensate at tree level, as magnetic field in
 the Higgs
model, and acts to restor symmetry. That was the reason why it has been
allowed for in lower order in Ref. \cite{Shap1}. But, as we will see for
strong fields and heavy $m_H$, the form of the EP curve in the broken
 phase is very sensitive to the change of the parameters. In the
restored phase, there is a number of terms having  order $ \sim
 (gH_Y)^{3/2} T$ which can influence the temperature of the phase
transition. So, to have an adequate picture of the phenomenon
investigated the radiation corrections in the field must be calculated
in both the broken and the restored phases. 

In the present paper the EW phase transition in the constant strong hypercharge
magnetic field is investigated within the consistent EP including the one-loop
and ring diagram contributions. All bosons and fermions are taken into account
with their actual masses (in particular, the t-quark mass is 
175 GeV). So, the only free parameter remains the mass $m_H$. We assume it to
be in the energy range 75 GeV $ \le m_H \le 115$ GeV, in order to take account
of the modern experemental low limit $m_H \ge$  90 GeV. We calculate the
contribution
of ring diagrams in the external fields. As it is well known, these diagrams
cancel the imaginary terms of the one-loop effective  potential and the total
potential is real at sufficiently high temperatures \cite{Tak}. Due to the
ring diagrams in the
field, there is also cancellation of an instability generated in strong
 magnetic fields in the
$W$-boson sector. This instability appears because of the presence  in the
$W$-boson  spectrum of
the tachyonic mode $\epsilon^2 = p^2_3  +  M^2_w - eH $ (see survey
\cite{Sk2}).
This mode is  the transversal one. So, to treat the problem carefully
the ring diagrams of the mode with the transversal effective mass at
nonzero $H, T$ have to be added. This requires calculation of the $W$-
boson mass operator at high temperature and strong fields. We present
 the relevant results also.  With such the term included the total EP is
real at sufficiently high temperatures and suitable to
investigate symmetry behaviour.

 For the bosonic part of the Salam-Weinberg model
  the phase transitions in magnetic fields at high temperature
have been studied in one-loop approximation in Refs. \cite{Cha}- \cite{Sk1}, \cite{AO},
\cite{MDT}, \cite{SVZ2}.  However,
the aspect of the EW phase transition, which was not investigated, 
 is the influence of the correlation corrections described by the ring
diagrams at high temperatures and strong fields.  At zero field it was
studied  in detail in Refs. \cite{Tak}, \cite{Car}, \cite{Din} but for
not heavy mass $m_t$.  In particular, in Ref. \cite{Car} $m_t$ was chosen of
order
 $ \sim 110$ GeV. So, for the present day experimental data it should be revised.

In what follows, considering EP in the broken phase we will write $H$ for the
 usual external magnetic field remembering that it equals to
$H = H_Y cos \theta$. The $Z$-component  of the field $H_Y$ is screened by
the scalar field
condensate $\phi_c$. The constant extenal field  is a good approximation for the
 description of the inital
stage of the first order EW phase transition when the bubles are not large, as
 it was discussed in Ref. \cite{Elm1}. 

The content is as follows. In  Sects. 2, 3 the one-loop contributions  of
bosons and fermions to the EP $V^{(1)}(T, H, \phi_c)$ are calculated in the
form convenient for numerical investigations. In Sect. 4 we compute the
contributions of ring diagrams. In Sect. 5 the EP for the restored phase is
calculated. In Sect. 6 symmetry behaviour at high temperatures and
strong external fields is investigated and it is shown  the EW
phase transition is of first order for the field strengths $H \sim 10^{22} -
10^{23}$G
but the baryogenesis condition is not satisfied. For stronger fields it
becomes of second order. Thus, we come to conclusion that in the
 SM  baryogenesis does not survive under smooth external hypermagnetic
field. The  comparison of our results with that of other approaches  and discussion  
 are given in Sects. 6, 7.

\section{Boson contributions to $V^{(1)}(T,H,\phi_c)$}

The Lagrangian of the boson sector of the Salam-Weinberg model is

\begin{eqnarray} \label{1}
 L  = -\frac{1}{4} F^a_{\mu\nu} F^{\mu\nu}_a
-\frac{1}{4} G_{\mu\nu} G^{\mu\nu} + (D_{\mu}\Phi)^+ (D^{\mu}\Phi)
\nonumber\\
+\frac{m^2}{2}(\Phi^+ \Phi) - \frac{\lambda}{4} (\Phi^+\Phi)^2,
\end{eqnarray}
where the standard notations are introduced
\begin{eqnarray}
F^a_{\mu\nu} = \partial_{\mu}A^a_{\nu}-\partial_{\nu}A^a_{\mu}
+ g\varepsilon^{abc} A^b_{\mu}A^c_{\nu},
\nonumber\\
G_{\mu\nu}  = \partial_{\mu}B_{\nu}-\partial_{\nu}B_{\mu},
\nonumber\\
D_{\mu}  =  \partial_{\mu} + \frac{1}{2}ig A^a_{\mu}\tau^a + \frac{1}{2}ig'
B_{\mu}.
\end{eqnarray}
\nonumber\\
The vacuum expectation value of the field $\Phi$ is
\begin{equation} \label{2} <\Phi> = \left(\begin{array}
{c}0 \\ \phi_c \end{array} \right).
\end{equation}
The fields corresponding to the $W$-, $Z$-bosons and photons, respectively, are
\begin{eqnarray}  W^{\pm}_{\mu} =\frac{1}{\sqrt{2}}(A^1_{\mu} \pm
iA^2_{\mu}),
\nonumber\\
 Z_{\mu} =\frac{1}{\sqrt{g^2 + g'^2}}(gA^3_{\mu} - g'B_{\mu}),
\nonumber\\
 A_{\mu} =\frac{1}{\sqrt{g^2 + g'^2}}(g'A^3_{\mu} + gB_{\mu}).
\end{eqnarray}
\nonumber\\
To incorporate an interaction with an external hypermagnetic field we add the term
$\frac{1}{2}
        \vec{H}\vec{H_Y}$ to the Lagrangian. The value of the macroscopic magnetic
        field generated inside the system will be determined by minimization of free
energy.  Interaction with a classical electromagnetic field is introduced as usually by
splitting the potential in
two parts: $A_{\mu}= \bar{A_{\mu}} + A^{R}_{\mu} $, where $A^{R}$
describes a
radiation field and $\bar{A} = (0,0,Hx^1,0)$ corresponds to the constant
magnetic field directed along the third axis. We make use of the gauge-fixing
conditions \cite{Sk2}
\begin{equation} \label{3} \partial_{\mu}W^{\pm \mu} \pm ie\bar{A_{\mu}}
W^{\pm \mu} \mp i\frac{g\phi_c}{2\xi}\phi^{\pm} = C^{\pm}(x),
\end{equation}
\begin{equation} \label{4} \partial_{\mu}Z^{\mu} - \frac{i}{\xi'}
(g^2 + g'^2)^{1/2}\phi_c\phi_{z} = C_z ,
\end{equation}
where $ e = g sin \theta, tang \theta = g'/g, \phi^{\pm}, \phi_{z}$ are the
Goldstone fields, $\xi, \xi' $ are the
gauge fixing parameters, $C^{\pm}, C_z$ are arbitrary functions and $\phi_c$ is
the value of the scalar condensate. In what follows, all calculations will be done in
the general relativistic renormalizable gauge (\ref{3}), (\ref{4}) and after
that we  set $\xi, \xi' = 0$ choosing the unitary gauge.

To compute the EP $V^{(1)}$ in the background magnetic field let us
introduce the proper time (s-representation) for the Green functions
\begin{equation} G^{ab}= - i \int\limits_{0}^{\infty} ds \exp(-is {G^{-1}}^{ab})
\end{equation}
\nonumber\\
and use the
method of Ref. \cite{Cab}, allowing in a natural way to incorporate the
temperature
into this formalism. A basic formula of Ref. \cite{Cab} connecting the
Matsubara-Green functions with the Green functions at zero temperature is needed,
\begin{equation} \label{5} G^{ab}_k(x, x';T) = \sum\limits_{-\infty}^{+\infty}
(-1)^{(n+[x])\sigma_k} G^{ab}_k(x-[x]\beta u, x'- n\beta u),
\end{equation}
where $G^{ab}_k$ is the corresponding function at $T=0, \beta =1/T, u = (0,0,0,1),$
the symbol $[x]$ means the integer part of $x_{4}/\beta, \sigma_k = 1$
in the case of physical fermions and $\sigma_{k} =0$ for the boson and the ghost fields.
The Green functions in the right-hand side of formula (\ref{5}) are the matrix
elements of the operators $G_k$ computed in the states $\mid x',a)$ at $T=0$,
and in the left-hand side the operators are averaged in the states with $T\not=
 0$. The corresponding functional spaces $U^{0}$ and $U^{T}$ are different but
in the limit of $T \rightarrow 0$ $ U^{T}$ transforms into $U^{0}$.

The one-loop contribution into EP is given by the expression
\begin{equation} \label{6} V^{(1)} = - \frac{1}{2} Tr\log G^{ab},
\end{equation}
where $G^{ab}$ stands for the propagators of all the quantum fields $W^{\pm},
\phi^{\pm},...$ in the background magnetic field $H$. In the s-representation
the calculation of the trace can be done in accordance with the formula \cite{Sch}
\begin{equation}\label{11}
 Tr\log G^{ab} = - \int\limits_{0}^{\infty} \frac{ds}{s}
tr \exp(-is G^{-1}_{ab} ).
\end{equation}
Details of calculations based on the s-representation and the formula (\ref{5})
can be found, for instance, in Refs. \cite{Cab}, \cite{Rez}, \cite{Sk3}. The terms
with $n=0$ in Eqs. (\ref{5}), (\ref{6}) give the zero temperature expressions for the
Green functions and the
effective potential $V^{(1)}$, respectively. They are the only terms possessing
divergences. To eliminate them and uniquely fix the potential we make use the
following renormalization conditions at $H, T = 0$  \cite{Rez}:
\begin{equation} \label{7} \frac{\partial^2 V(\phi,H)}{\partial H^2}\mid_{H=0,
\phi=\delta(0)} = \frac{1}{2},
\end{equation}
\begin{equation} \label{8} \frac{\partial V(\phi,H)}{\partial \phi}\mid_{H=0,
\phi=\delta(0)} = 0,
\end{equation}
\begin{equation} \label{9} \frac{\partial^2 V(\phi,H)}{\partial \phi^2}
\mid_{H=0,\phi=\delta(0)} = \mid m^2 \mid,
\end{equation}
where $V(\phi,H)=V^{(0)}+V^{(1)}+ \cdots$ is the expansion with respect to the number of
loops
and $\delta(0)$ is the vacuum value of the scalar field determined in tree
approximation.

It is convenient for what follows to introduce dimensionless quantities:
$h=H/H_0 ~(H_0=M^2_w/e), \phi=\phi_c/\delta(0), K =m_H^2/M_w^2,$ $B=\beta
M_w, \tau=1/B = T/M_w,$$ {\cal V}= V/H^2_0$ and $M_w = \frac{g}{2}\delta(0)$.

After the reparametrization the scalar field potential is explicitely expressed in
terms of the ratio $K,$
\begin{equation} \label{10} {\cal V}^{(0)} = \frac{h^2}{2} + K sin^2 \theta (- \phi^2
 + \frac{\phi^4}{2} ).
\end{equation}
Remind that $h$ is the «electromagnetic» component of the hypercharge field $h_Y$
which is
unscreened in the broken phase. In the restored phase it will be convenient to work in
terms of the initial fields and we will carry out the corresponding calculations
later.

The renormalized one-loop EP is given by the sum of the functions
\begin{equation} \label{11}
{\cal V}_1 = {\cal V}^{(0)} + {\cal V}^{(1)}(\phi,h,K)
 + \omega^{(1)}(\phi,h,K,\tau),
\end{equation}
where ${\cal V}^{(1)}$ is the one-loop EP at $T=0$, which has been studied
already in Ref. \cite{Sk2}. It has the form:
\begin{equation} \label{12} {\cal V}^{(1)} = {\cal V}^{(1)}_{w,z} +
{\cal V}^{(1)}_{\phi},
\end{equation}
where
\begin{eqnarray} \label{12a} {\cal V}^{(1)}_{w,z} = &\frac{3\alpha}{\pi}&
[h^2 log \Gamma_1 (\frac{1}{2} +\frac{\phi^2}{2h}) + h^2 \zeta^{'}(-1) +
\frac{1}{16} \phi^4 - \frac{1}{8} \phi^4 log\frac{\phi^2}{2h} + \frac{1}{24}
h^2 \nonumber\\  &-& \frac{1}{24} h^2 log(2h)]
+ \frac{\alpha}{2\pi} [ - 2h^2 + (h^2 + h \phi^2) log(h + \phi^2)\nonumber\\
&+& (h^2 - h\phi^2 ) log \mid h - \phi^2 \mid ]
+ i \frac{1}{2} \alpha h (\phi^2 - h) \theta (h - \phi^2),
\end{eqnarray}

\begin{eqnarray} \label{12b} {\cal V}^{(1)}_{\phi} = &K& sin^2 \theta
_w ( - \phi^2 + \frac{1}{2} \phi^4 )  \nonumber\\
&+&  \frac{3 \alpha}{4\pi}( 1 + \frac{1}{2 cos^4 \theta}) (\frac{1}{2} \phi^4
log \phi^2 - \frac{3}{4} \phi^4 + \phi^2 )  \nonumber\\
&+& \frac{\alpha K^2}{32 \pi} [(\frac{9}{2} \phi^4 - 3 \phi^2 +
\frac{1}{2} ) log \mid \frac{3 \phi^2 - 1}{2} \mid - \frac{27}{4}\phi^4 +
\frac{21}{2} \phi^2 ]
\end{eqnarray}
and $\omega^{(1)}$ is the temperature dependent contribution to the EP
determined by the corresponding terms of formulae (\ref{5}), (\ref{6}) with $n \not=
0$.

We outline the used calculation procedure considering the $W$-boson
contribution as an example \cite{Sk3},
\begin {eqnarray} \label {13}
\omega^{(1)}_{w} =  \frac{\alpha}{2\pi}
\int\limits_{0}^{\infty}\ \frac{ds}{s^2}\ e^{-is(\phi^2/h)} \Bigl[\frac{1 +
2 \cos 2s}{\sin s} \Bigr]
\sum\limits_{1}^{\infty} \exp(ihB^2 n^2/4s).
\end{eqnarray}
As Eq. (\ref{12a}), this expression contains an imaginary part for $h > \phi^2$
appearing due to the tachyonic mode $ \varepsilon^2 = p^2_3 + M^2_w - eH $ in
the $W$-boson spectrum \cite{Sk2}, \cite{Niol}, \cite{Sk78}. It can be explicitly
calculated by means of an
analytic continuation taking into account the shift $s \rightarrow s -i0 $ in the  
$s$-plane. Fixing $\phi^2/h > 1$ one can rotate clockwise the integration contour
in the $s$-plane and direct it along the negative imaginary axis. Then, using
the identity
\begin{equation} \frac{1}{\sinh s} = 2 \sum\limits_{p=0}^{\infty} e^{-s(2p+1)},
\end{equation}
\nonumber\\
and integrating over $s$ in accordance with the standard formula
\begin{equation} \label{14} \int\limits_{0}^{\infty} ds s^{n-1} \exp(-\frac{b}
{s} - as) = 2 (\frac{b}{a})^{n/2} K_{n}(2\sqrt{ab}),
\end{equation}
$(a,b > 0)$, one can represent the expression (\ref{13}) in the form
\begin{equation} \label{14a}  Re \omega^{(1)}_w = - 4 \frac{\alpha}{\pi}
\frac{h}{B} ( 3\omega_0 + \omega_1 - \omega_2 ),
\end{equation}
where
\begin{equation} \omega_0 = \sum\limits_{p=0}^{\infty} \sum\limits_{n=1}^
{\infty} \frac{x_p}{n} K_1(nBx_p) ;~ x_p = (\phi^2 + h +2ph )^{1/2},
\end{equation}
\nonumber\\
\begin{equation} \omega_1 = \sum\limits_{n=1}^{\infty} \frac{y}{n} K_1(nBy),~
y = (\phi^2 - h )^{1/2}.
\end{equation}
\nonumber\\
We have in the range of parameters $ \phi^2 < h $ after analytic continuation
\begin{equation} \omega_1 = -\frac{\pi}{2} \sum\limits_{n=1}^{\infty}
\frac{\mid y \mid}{n} Y_1(nB\mid y \mid) ,
\end{equation}
\nonumber\\
\begin{equation} \omega_2 = \sum\limits_{n=1}^{\infty} \frac{z}{n} K_1(nBz),
z = (\phi^2 + h )^{1/2},
\end {equation}
 $K_n(x), Y_n(x)$ are the Bessel functions. The
imaginary part of $\omega^{(1)}_w$ is given by the expression
\begin{equation} \label{15} Im \omega_1= -2\alpha\frac{h}{B}\sum\limits_{n=1}
^{\infty} \frac{\mid y \mid }{n} J_1(nB\mid y \mid),
\end{equation}
$J_1(x)$ is the Bessel function. As it is well known, the imaginary part of EP is
signaling the instability of the system. In what follows we shall
consider mainly symmetry behaviour described by the real part of the EP.
As the imaginary part,  it will be cancelled  in the consistent calculation
including the one-loop and ring diagram contributions to the EP.

Carrying out similar calculations for the $Z$- and Higgs bosons, we obtain
\cite{Rez}:
\begin{equation} \label{16} \omega_z = - 6\frac{\alpha}{\pi} \sum\limits_{n=
1}^{\infty} \frac{\phi^2}{\cos^2 \theta_w n^2 B^2}
K_2(\frac{nB\phi}{\cos\theta}),
\end{equation}
\begin{equation} \label{17}
Re \omega_{\phi} = \Bigl\{
\begin{array}{c}-2 \frac{\alpha}{\pi} \sum\frac{t^2}{B^2 n^2} K_2(nBt)\\
\alpha \sum\limits_{n=1}^{\infty} \frac{\mid t \mid^2}{n^2 B^2} Y_2(nB\mid t
\mid)\end{array}\Bigr\},
\end{equation}
where the variable $t = [K_w(\frac{3\phi^2-1)}{2})]^{1/2}$ at $3\phi^2>1$ and
series with the function $Y_2(x)$ has to be calculated  at $3\phi^2<1$.
The
corresponding imaginary term is also cancelled as it will  be  shown below.

The above expressions (\ref{12}), (\ref{14a}), (\ref{16}), (\ref{17}) will be
used in the numerical studying of symmetry behaviour at different $H, T$.
There is a cancellation of a number of terms from the zero-temperature
contributions given Eqs. (\ref{12}) and $T$-depended ones. This fact has a
general character and was used in checking of the correctness of calculations.

\section{Fermion contributions to $V^{(1)}(H,T,\phi_c)$}

To find the convenient form of the fermion contribution to the EP
let us consider the standard  unrenormalized  expression  written in the $s$ 
-representation \cite{Elm}:
\begin{equation} \label{18} V^{(1)}_{f} =  \frac{1}{8\pi^2}\sum\limits_
{n=-\infty}^{\infty} (-1)^n \int\limits_{0}^{+\infty}\frac{ds}{s^3}
e^{-(m^2_fs + \beta^2n^2/4s)} (eHs)coth (eHs) ,
\end{equation}
$m_f$ is a fermion mass. Here, we have incorporated the equation (\ref{5}) to
introduce the temperature dependence. In what follows, we shall take into account
the contributions of all fermions - leptons and quarks - with the values of masses
equalled to the present day data. Usually, considering  symmetry behaviour without external fields
one restricts themself by the $t$-quark  contribution, only. But in the case of the
external field applied this is not a good idea, since the dependence of $V^{(1)
}$ on $H$ is the complicate function of the parameters $m^2_{f},~ eH,~ T$. At some
fixed values of $H, T$ fermions with the definite corresponding masses are dominant.
 For instance, at high  temperature the liding term of
$V^{(1)}_f$ is $\sim H^2 log \frac{T}{m_f}$. Hence it follows that light  fermions
are important . In general,  a very complicate dependence on the field
takes place. We include this in the total, carrying out  numerical
calculations and summing up over all the fermions. Now, separating the zero
temperature terms by  means of the relation $\sum\limits_{-\infty}^
{+\infty} = 1 + 2 \sum\limits_{1}^{\infty}$ and introducing the parameter
$K_{f}=m^2_{f}/M^2_w = G^2_{Yukawa}/g^2$, we obtain for the  zero
temperature
fermion contribution to the dimensionless EP,
\begin{eqnarray} \label{19}
{\cal V}_{f}(h,\phi) & =&\frac{\alpha}{4\pi} \sum\limits_{f} K^2_{f}(- 2 \phi^2
+ \frac{3}{2} \phi^4 - \phi^4 log \phi^2 )\nonumber\\
 &-&\frac{\alpha}{\pi}\sum\limits_{f}(q^2_f\frac{h^2}{6}
\log\frac{2\mid q_f \mid h}{K_f})
\nonumber\\
&-& \frac{\alpha}{\pi} \sum\limits_{f} \Bigl.[2q^2_f h^2 \log\Gamma_1(
\frac{K_f\phi^2}{2\mid q_f\mid h}) + (2\zeta'(-1)-\frac{1}{6})q^2_f h^2
\nonumber\\
&+&\frac{1}{8} K^2_f\phi^4 + (\frac{1}{4}K^2_f\phi^4 - \frac{1}{2}K_f\mid
q_f
\mid h\phi^2) \log\frac{2\mid q_f \mid h}{K_f \phi^2}\Bigr],
\end{eqnarray}
where $q_f$ is a fermion electric charge, the sum $\sum\limits_{f} = \sum
\limits_{f=1}^{3}(leptons) + 3 \sum\limits_{f=1}^{3}(quarks)$ counts the
contributions of leptons and quarks with their electric charges. The
 function $\Gamma_1$  is defined as follows \cite{Ditt} (see also survey \cite{Sk2}):
\begin{equation} \log\Gamma_1(x) = \int\limits_{0}^{x} dy \log\Gamma(y) +
\frac{1}{2}x(x-1) - \frac{1}{2}x\log(2\pi).
\end{equation}
\nonumber\\

The finite temperature part can be calculated in a way described in
the previous section. In the dimensionless variables it looks as follows:
\begin{eqnarray} \label{20} \omega_{f}&=& 4 \frac{\alpha}{\pi}\sum
\limits_{f}\Bigl\{\sum_{p=0}^{\infty}\sum_{n=1}^{\infty}(-1)^n \Bigl[\frac{
(2ph + K_f\phi^2)^{1/2} h}{Bn} K_1((2ph + K_f\phi^2)^{1/2}Bn)
\nonumber\\
&+& \frac{((2p+2)h + K_f\phi^2)^{1/2}}{Bn} h
K_1(((2p+2)h + K_f\phi^2)^{1/2}Bn)\Bigr]\Bigr\}
\end{eqnarray}
Again, a number of terms from Eqs. (\ref{19}) and (\ref{20}) are cancelled being
summed up, as in the bosonic sector.

These two expressions and the boson contributions obtained in Sect. 2 will be
used in numerical investigations of symmetry behaviour.

\section{Contribution of ring diagrams}

It was shown by Carrington \cite{Car} that at $T \not = 0$ the consistent
calculation
of the EP based on generalized propagators, which include the polarization
operator insertions, requires that ring diagrams have to be added simultaneously
with the one-loop part. These diagrams  essentially affect
the phase transition at high temperature and zero field \cite{Tak}, \cite{Car},
\cite{Din}.
 Their importance at $ T$ and $H \not= 0$ was also pointed out in literature
 \cite{SVZ1}, \cite{SVZ2} but, as far as we know, this part of the EP has   not been
calculated, yet.

As it is known \cite{Kal}, \cite{Tak}, the sum of ring diagrams describes a dominant
contribution
of large distances. It  non-negligibly differs from zero only in the case when
massless states
appear in a system. So, this type of diagrams has to be calculated when a
symmetry restoration is investigated. To find the correction $V_{ring}(H, T)$ at
high temperature and magnetic field the polarization operators of the Higgs
particle, photon and $Z$-boson at the considered background have to be
computed. Just these calculations have been announced in Refs. \cite{SVZ1},
\cite{SVZ2}. Then, $V_{ring}(H, T)$ is given by series depicted in figures 1, 2.
\begin{figure}
\begin{center}
  \epsfxsize=0.6637222\textwidth
  \epsfbox[0 250 612 352]{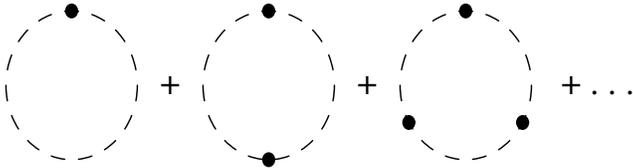}
  \caption{The Higgs field ring diagrams giving contribution to the effective
potential. Blobs stand for the neutral scalar field polarization operator calculated at zero
momentum.}
\end{center}
\end{figure}
\begin{figure}
\begin{center}
  \epsfxsize=0.6637222\textwidth
  \epsfbox[0 250 612 352]{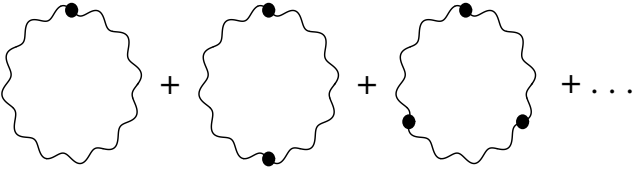}
  \caption{The photon and Z-boson ring diagrams giving contribution to the effective
potential. Blobs stand for the polarization operators of the fields calculated at zero
momenta. }
\end{center}
\end{figure}
Here, a dashed line describes the Higgs particles, the wavy lines represent photons
and $Z$-bosons, the blobs represent the polarization operators in the limit of
zero momenta. As it is also known \cite{Car}, in order to calculate the
contribution
of ring diagrams not the total polarization operators $\Pi_{\mu\nu}(k, T, H)$
but only their limits at zero momenta, $\Pi_{00}(k=0, T, H)$, are
sufficient. This limit, called the Debye mass,  can be calculated  from the EP of
the special type. This fact considerably simplifies our task.

Now, let us turn to calculations of $V_{ring}(H, T)$. It is given by the standard
expression \cite{Tak}, \cite{Car}, \cite{SVZ1}:
\begin{equation} \label{21}
V_{ring} = - \frac{1}{12\pi\beta} Tr\{[M^2(\phi) +
\Pi_{00}(0)]^{3/2} - M^3(\phi)\},
\end{equation}
where trace means the summation over all the contributing states, $M(\phi)$ is
the tree mass of the corresponding state. The functions  $\Pi_{00}(0)$ are:
 $\Pi_{00}(0)$ =$ \Pi(k=0,T,H)$ for the
Higgs particle; $\Pi_{00}(0)=\Pi_{00}(k=0,T,H)$ - the zero-zero
components
of the polarization functions of gauge fields in the magnetic field taken at zero
momenta.
The above contributions are of order $ \sim g^3 (\lambda^{3/2}) $ in the coupling
constants whereas the two-loop terms are to be of order $\sim g^4 (\lambda^{2})$. For
$\Pi_{00}(0)$ the high temperature limits of polarization functions have
to be substituted which have  order $\sim T^2$ for leading terms and
$\sim g\phi_c T, (gH)^{1/2}T (\phi_c/T << 1 , (gH)^{1/2}/T << 1)$ for subleading
ones.

For the next step of calculation, we remind that the effective potential is
the generating functional of the one-particle irreducible Green functions at
zero external momenta. So, to have $\Pi(0)$ we can just calculate the second
derivative with respect to $\phi$ of the potential $V^{(1)}(H,T,\phi)$ in the
limit of high temperature, $ T >>\phi , T >> (eH)^{1/2}$, and then set $\phi = 0$.
This
limit can be obtained by means of the Mellin
transformation technique (see, for instance, \cite{Sk3}) and the result looks as
follows:
\begin{eqnarray} \label{22} V^{(1)}(H,\phi,T)_{\mid_{T\rightarrow \infty}} &=& \left.[\Bigl(
\frac{C_f}{6}\phi^2_c + \frac{\alpha\pi}{2 cos^2\theta_w}\phi^2_c + \frac{g^2}
{16}\phi^2_c \Bigr) T^2  \right.]
\nonumber\\
&+& \left.[ \frac{\alpha \pi}{6} (3\lambda\phi^2_c - \delta^2(0))T^2 - \frac{
\alpha}{cos^3\theta} \phi^3 T - \frac{\alpha}{3} (\frac{3\lambda\phi^2_c -
\delta^2(0)}{2})^{3/2} T \right.]
\nonumber\\
&-& \frac{1}{2\pi} (\frac{1}{4}\phi^2_c + gH)^{3/2} T + \frac{1}{4\pi} eH T
(\frac{1}{4}\phi^2_c + eH )^{1/2}
\nonumber\\
 &+& \frac{1}{2\pi} eH T (\frac{1}{4} \phi^2_c -
eH )^{1/2}.
\end{eqnarray}
The parameter $C_f = \sum\limits_{i=1}^{3} G^2_{il} + 3\sum\limits_{i=1}^{3}
G^2_{iq}$ determines the fermion contribution of  order $\sim T^2$ having
relevance to our problem. We also have omitted $\sim T^4$ contributions to the EP.
The terms  of the type $\sim log [T/f(\phi,H)]$ cancel the logarithmic terms in
the temperature independent parts (\ref{11}), (\ref{18}). Considering
the high temperature limit we restrict ourselves to linear and quadratic in $T$
terms, only.

The one else important expression, which also should be taken into account, is   the
linear
in $H$ term of the zero temperature EP Eq. (\ref{19}), which looks as follows:
\begin{equation} \label{lh}
V^{(1)}_{f,l}(H,\phi_c)/H_0^2 = - \frac{\alpha}{2\pi} \phi^2 \sum\limits_f K_f \mid q_f H \mid.
\end{equation}
It significantly influences symmetry behaviour and contributes to the Debye mass in
strong
fields.

Now, differentiating these expressions twice with respect to $\phi$ and setting
 $\phi=0$, we obtain
\begin{eqnarray} \label{23} \Pi_{\phi}(0) &=& \frac{\partial^2 V^{(1)}
(\phi,H,T)}{\partial \phi^2} \mid_{\phi=0}
\nonumber\\
&=& \frac{1}{24 \beta^2}\Bigl( 6\lambda + \frac{6 e^2}{\sin^2 2\theta_w}
+ \frac{3 e^2}{\sin^2 \theta_w} \Bigr) \nonumber\\
&+& \frac{2\alpha}{\pi} \sum\limits_{f}\Bigl[ \frac{\pi^2 K_f}{3\beta^2} - \mid q_f H \mid K_f  \Bigr]
\nonumber\\
&+& \frac{(eH)^{1/2}}{8\pi \sin^2\theta_{w}\beta} e^2 (3\sqrt{2} \zeta(-\frac{1}{2},\frac{1}{2}) - 1 ).
\end{eqnarray}
The terms $\sim T^2$ in Eq. (\ref{23}) give standard contributions to
temperature mass squared coming from the boson and the fermion sectors.
The $H$-dependent term is negative since the difference in the brackets is
$3\sqrt{2}\zeta (-\frac{1}{2},\frac{1}{2}) - 1 \simeq - 0,39$. Formally, this
may result in the negativeness of the $\Pi(0)_{\phi}$ for very strong fields
$(eH)^{1/2} > T $. But this happens in the range of parameters where asymptotic
axpansion is not applicable. Substituting expression (\ref{23}) into Eq. (\ref{21})
we  obtain (in the dimensionless variables)
\begin{equation} \label{24} {\cal V}^{\phi}_{ring} = - \frac{\alpha}{3B}
\Bigl\{(\frac{3\phi^2 - 1}{2} K  + \Pi_{\phi}(0) \Bigr\}^{3/2} + \frac{\alpha}
{3B} K (\frac{3 \phi^2 - 1}{2})^{3/2}.
\end{equation}
As one can see, the last term of this expression cancels  the  fourth term in the Eq. 
(\ref
{22}), which becomes imaginary at $3\phi^2 < 1$. This is the  important
 cancellation preventing  the infrared instability  at high temperature.

Before we proceed, let us note that  Eq. (\ref{22}) contains the other term (the last
one)
which becomes imaginary  for strong  magnetic fields or small $\phi^2$.  It reflects the
known instability  in the $W$-boson spectrum which is discussed for many years in
literature (see  papers \cite{Rez}, \cite{SVZ1}, \cite{SVZ2}, \cite{Sk3} and
references
therein).  But it also  will be cancelled out when the contribution of  ring diagrams
with
the  unstable mode  is added.

To find the $H$-dependent Debye masses of photons and $Z$-bosons the
following procedure will be used. We calculate the one-loop EP of the
$W$-bosons and fermions in a magnetic field and some "chemical
potential", $\mu,$  which plays the role of  an auxiliary parameter.
Then, by differentiating them twice with respect to $\mu$ and setting $\mu = 0$
the mass squared $m^2_D$ will be obtained. Let us first demonstrate that in more
detail for the case of fermion contributions where the result is known.

The temperature dependent part of the one-loop EP of constant magnetic field
at a non-zero chemical potential induced by an electron-positron vacuum
polarization is \cite{Elm}:
\begin{equation} \label{25} V^{(1)}_{ferm.} =
  \frac{1}{4\pi^2} \sum\limits_{l=1}^{\infty}
(-1)^{l+1}\int\limits_{0}^{\infty} \frac{ds}{s^3} exp(\frac{-\beta^2 l^2}
{4s} - m^2s ) (eHs) coth(eHs) cosh(\beta l\mu),
\end{equation}
where $m$ is the electron mass, $e = g sin \theta$ is the electric charge and the
proper-time representation is used. Its second derivative with respect to $\mu$
taken at $\mu = 0$ can be written in the form,
\begin{equation} \label{26}  \frac{\partial^2 V^{(1)}_{ferm.}}{\partial \mu^2}=
\frac{eH}{\pi^2}\beta^2 \frac{\partial}{\beta^2}\sum\limits_{l=1}^{\infty}
(-1)^{l+1}\int\limits_{0}^{\infty} \frac{ds}{s} exp(-m^2s -\beta^2 l^2/4s)
coth(eHs).
\end{equation}
Expanding $coth (eHs) $ in series and integrating over $s$ in accordance with
formula (\ref{14}) we obtain in the limit of $ T >> m, T >> (eH)^{1/2}$:
\begin{equation} \label{27} \sum\limits_{l=1}^{\infty} (-1)^{l+1} [\frac{8m}
{\beta l} K_1(m\beta l) + \frac{2}{3}\frac{(eH)^2 l\beta}{m} K_1(m\beta l)+
\cdots ].
\end{equation}
The series in $l$ can easily be calculated by means of the Mellin transformation
(see Refs.\cite{Sk3}, \cite{SVZ2}). To have the Debye mass squared it is necessary
to differentiate Eq. (\ref{26}) with respect to $\beta^2$ and to take into
account the relation of  the parameter $\mu$ with the zero component of the
electromagnetic potential: $\mu \rightarrow  ieA_0$ \cite{SVZ1}. In this way we
obtain finally,
\begin{equation} \label{28} m^2_{D} = g^2 sin^2\theta \Bigl[ \frac{T^2}{3} -
\frac{1}{2\pi^2} m^2 + O((m\beta)^2, (eH\beta^2) ) \Bigr].
\end{equation}
This is the well known result calculated from the photon polarization operator
 \cite{VZM}. As  one can see, the dependence on $ H$ appears in the order
$\sim T^{-2}$. To find the total fermion contribution to $m^2_D$ one should
sum up the
expression (\ref{28}) over all fermions and substitute their electric charges.

To find $m^2_D$ for $Z$-bosons it is sufficient to allow for the
fermion coupling to the $Z$-field. It can be done by substituting $ \mu \rightarrow
i(g/2 cos \theta + g sin^2 \theta ) $ and the result differs from Eq.
(\ref{28}) by the coefficient at the brackets in the right-hand side which
has to be replaced, $g^2 sin^2 \theta \rightarrow g^2 (\frac{1}{4 cos^2 \theta} +
tang^2 \theta) $. One also should add the terms coming due to the neutral
currents and the part of fermion-Z-boson interaction which is not
reproduced by the above substitution:
\begin{equation} m^{2'}_{D} = \frac{g^2}{8 cos^2 \theta} (1 + 4 sin^4
\theta) T^2 .
\end{equation}
\nonumber\\

Now, let us apply this procedure to the case of the $W$-boson contribution.
The corresponding EP (temperature dependent part) calculated at non-zero
$T, \mu
$ in the unitary gauge looks as follows,
\begin{eqnarray} \label{29} V^{(1)}_w &=& - \frac{eH}{8\pi^2}\sum\limits_{l=1}^{
\infty}\int\limits_{0}^{\infty} \frac{ds}{s^2}exp(-m^2 s -l^2\beta^2/4s)\\ \nonumber
&&[\frac{3}{sinh(eHs)}+ 4 sinh(eHs)] cosh(\beta l\mu).
\end{eqnarray}
All the notations are obvious. The first term in the squared brackets gives
the triple contribution of the charged scalar field and the second one is due
to the interaction with a $W$-boson magnetic moment. Again, after
differentiation twice with respect to $\mu$ and setting $\mu = 0$ it can be
written as
\begin{equation} \label{30} \frac{\partial^2 V^{(1)}_w}{\partial \mu^2}_{\mid
\mu =0} =
\frac{eH}{2\pi^2}\beta^2\frac{\partial}{\partial\beta^2} \sum\limits_{l=1}^
{\infty}\int\limits_{0}^{\infty}\frac{ds}{
s}exp(- \frac{m^2s}{eH} - \frac{l^2\beta^2eH}{4s})[\frac{3}{sinh(s)} + 4
sinh(s)].
\end{equation}
Expanding $sinh^{-1}s$ in a series over Bernoulli's polynomials,
\begin{equation} \frac{1}{sinh s} = \frac{e^{-s}}{s} \sum\limits_{k=0}^{\infty}
\frac{B_k}{k!}(-2s)^k,
\end{equation}
\nonumber\\
and carrying out all the calculations described above, we obtain for the $W$-
boson contribution to $m^2_D$ of the electromagnetic field,
\begin{eqnarray} \label{31} m^2_D &=& 3 g^2 sin^2 \theta [ \frac{1}{3} T^2 -
\frac{1}{2\pi} T(m^2 + g sin\theta H)^{1/2} - \frac{1}{8\pi^2}(g sin\theta H)
\nonumber\\
&+& O(m^2/T^2, (g sin\theta H /T^2)^2)].
\end{eqnarray}
Hence it follows that   spin does not affect the Debye mass in leading
order. Other interesting feature is that the next-to-leading terms are negative.

The contribution of the $W$-boson sector to the $Z$-boson mass $m^2_D$ is
given by the expression (\ref{31}) with the replacement $g^2 sin^2 \theta
\rightarrow g^2 cos^2 \theta$.

Substituting the expressions (\ref{28}) and (\ref{31}) into
Eq. (\ref{21}), we obtain the photon part $V^{\gamma}_{ring}$, where it is
necessary to express masses in terms of the vacuum value of the scalar
condensate $\phi_{c}$. In the same way the ring diagrams of $Z$-bosons $V^{z}_{
ring}$ can be calculated. The only difference is the mass term of $Z$-
field and the additional term in the Debye mass due to  the neutral current
$\sim \bar{\nu}\gamma_{\mu}\nu Z_{\mu}$. These three fields - $\phi, \gamma,
 Z$,- which become massless in the
restored phase, contribute into $V_{ring}(H, T)$ in the presence of the magnetic
field. At zero field, there are also terms due to the $W$-boson loops
in Figs. 1, 2 . But when $H \not = 0$ the charged  particles acquire masses $\sim eH$
and can be neglected.

  In the restored phase, the $W$-bosons  do not interact with
the hypermagnetic field and therefore give no field dependent
contributions.  

A separate consideration should be spared to the tachyonic (unstable)  mode in the
$W$-boson spectrum: $p^2_0 = p^2_3 + M^2_w - eH$.  First  of all, we  notice that
this mode is excited due to a spin interaction and it does not influence the
$G_{00}(k)$ component of the $W$-boson propagator.  Secondly,  in the fields
$eH \sim M^2_w$ the  mode becomes a long range state. Therefore,  it should be
included in  $V_{ring}(H,T)$ side by side  with the other considered neutral fields.  But
in
this case
it is impossible  to take advantage of formula (\ref{21}) and one has to return to the initial
EP containing the generalized propagators.

For our purpose  it will be convenient to make use of the  generalized
EP
written as the sum over the modes in the external magnetic field \cite{SVZ1},
\cite{SVZ2}:
\begin{equation} \label{TDVZ}  V^{(1)}_{gen} =  \frac{eH}{2\pi \beta} \sum
\limits_{l= -\infty}^{+ \infty} \int\limits_{- \infty}^{+ \infty} \frac{dp_3}{2\pi}
\sum\limits_{n = 0, \sigma = 0,\pm 1}^{\infty} log [\beta^2(\omega^2_l + \epsilon
^2_{n,\sigma,p_3} + \Pi(T,H) )] ,
\end{equation}
where $\omega_l = \frac{2\pi l}{\beta}$ ,  $\epsilon^2_n = p_3^2 + M^2_w + (2n + 1 -
2\sigma) eH $ and $\Pi(H,T)$ is the radiation mass squared of  $W$-bosons  in a magnetic
field at finite temperature.
Denoting   as $D^{- 1}_0(p_3,H.T)$ the sum $ \omega^2_l + \epsilon^2$, one can
rewrite eq. (\ref{TDVZ}) as follows:

\begin{eqnarray} \label{TDVZ1}
V^{(1)}_{gen} &= & \frac{eH}{2\pi\beta}
\sum\limits_{l=-\infty}^{+\infty}\int\limits_{-\infty}^{+\infty} \frac{d p_3}{2\pi}
\sum\limits_{n,\sigma} log[\beta^2 D^{- 1}_0(p_3,H,T)]
\nonumber\\
&+&  \frac{eH}{2\pi\beta} \sum\limits_{l = - \infty}^{+ \infty}  \int\limits_{-
\infty}^{+\infty} \frac{d p_3}{2\pi} \{ log[ 1 + ( \omega^2_l + p^2_3 + M^2_w - eH)^{-
1} \Pi(H,T) ] \nonumber\\
&+& \sum\limits_{n \not = 0, \sigma \not = +1 } log[ 1 +  D_0 ( \epsilon^2_n , H,T)
\Pi(H,T) ] \}.
\end{eqnarray}
Here,  the first term  is just the one-loop contribution of $W$-bosons,  the second one
gives the  sum of ring diagrams  of the unstable mode (as it can easily be verified by
expanding the logarithm into a series). The last term describes the  sum of the short
range modes in the magnetic field and should be omitted.

Thus, to get  $V^{unstable}_{ring}$ one has to compute the second term in Eq.  
(\ref{TDVZ1}). In the high temperature limit we obtain:
\begin{equation} \label {Vunst} V^{unstable}_{ring} = \frac{eH}{2\pi\beta} \{ (
M^2_w - eH + \Pi (H, T) )^{1/2} - ( M^2_w - eH )^{1/2} \}.
\end{equation}
By summing up  the one-loop EP and   all the  terms $ V_{ring}$, we arrive at the total
 consistent in leading order EP.

Let us mention the most important features of the above expression. It is seen that the
last
term in Eq. (\ref{Vunst}) exactly cancels the "dangerous" term in Eq. (\ref{22}). So,
 the EP is real and no instabilities appear at sufficently high temperatures when $\Pi
( H,T) > M^2_w - eH $.  To make a quantitative estimate of the range of validity of the
total EP
it is necessary to calculate the $W$-boson mass operator in a magnetic field at finite
temperature and hence to find $ \Pi (H,T)$. This is a separate and enough cogent
problem which is considered in  detail in a separate publication. Here, we only adduce
the result
of $\Pi (H, T)$ calculations \cite{SS}:
\begin{eqnarray} \label{munst}
\Pi_{unstable}(H, T) &=&  <n=0,\sigma= + 1\mid \Pi^{charged}_{\mu \nu} \mid n=0,
\sigma = + 1>
\nonumber \\
 &=&  \alpha [26,96 (eH)^{1/2} T  +  i 4 (eH)^{1/2}T] ,
\end{eqnarray}
where  the average value of the mass operator in the ground state of the $W$-boson
spectrum $ \mid n =0, \sigma = + 1> $  was taken. This expression has been computed
in the limit $eH/ T^2 << 1, B = M_w(H, T)/T << 1$, which is a good approximation since,
as it
will be shown below,  typical inverse temperatures for the symmetry restoration  are $B \sim
0.1 - 0.3$. Side by side with the real part responsible for the radiation mass squared
the expression
(\ref{munst}) contains the imaginary one describing the decay of the state. Its value is small
as  compare to the real part and  of  order of the  usual damping constant at high
temperature. So, Im $ \Pi (H,T)$ can be ignored in our problem. The radiation mass  squared
is positive and  acts  to stabilize the spectrum.  At $H = 0$ no screening is produced in
one-loop order, as it should be at finite temperatures for transversal modes \cite{Kal}.
Thus, we see that at high temperatures the effective $W$-boson  mass
squared, $(M^2_w)
_{eff.} = M^2_w - eH + \Pi (H,T),$ is positive. Therefore, no conditions for $W$-boson
condensation discussed in Refs. \cite{Sk1}, \cite{AO} are realized.
With this result obtained we conclude that our EP is real for temperatures corresponding to
the phase transition epoch.$^{*)}$
\footnote{  Expression  (\ref{munst}) disagrees with the corresponding  one of
Ref. \cite{Elp} where the average value of the gluon polarization operator in an abelian
chromomagnetic field was calculated in a weak field approximation and $\Pi(H,T)$ has been found to be zero.
Most probably,  the discrepancy is the concequence of the calculation procedure adopted by
these authors when the gluon polarization operator was computed at zero external field and
then its average value has been calculated in the state $\mid n = 0, \sigma = + 1> $ . Our
expression is the high temperature limit of the mass operator which takes account of the
external field exactly.}

\section{Effective potential of the restored phase}

Having obtained the EP at $\phi \not  = 0$ we are able to investigate the form of its
curve in the broken phase and determine the type of the EW phase transition for different
$m_H, h$. To describe more precise the restored phase one has also to calculate
radiation corrections to the external hypermagnetic field $H_Y$ at high temperature. Before
doing
that let us remind that at $\phi = 0$ the field $H_Y$  is completely unscreened. This
 means that in a covariant derivative describing
interaction with the external field one should include the $U(1)_Y$ term:
$D_{\mu} = \partial_{\mu} + \frac{1}{2} g' B_{\mu}^{ext}$. We set the potential
as before, $B_{\mu}^{ext} = ( 0, 0, H_Y, 0 )$.

In the restored phase $W$-bosons do not interact with $H_Y $.  The field dependent part
of the EP $V ( \phi = 0, H_Y, T )$ is non-zero due to the contributions of fermions and
scalars. However, the fermion part depends logarithmically on temperature ($ \sim
\frac{(g'/2)^{2}}{4 \pi} H^2_Y log T/ T_0 $) and can be neglected as compared to the tree
level term $\frac{1}{2} H_Y^2$.  The scalar field contribution to  the one-loop EP is
\begin{eqnarray} \label{Yscal}
V^{(1)}_{sc}(H_Y, T ) =  &-&\frac{(g'/2)^2 H_Y^2}{24 \pi^2} ln (T/T_0)\nonumber\\
 &+&
\frac{((g'/2) H_Y)^{3/2} T}{ 6 \pi} + O (1/T).
\end{eqnarray}
The term logarithmically dependent on $T$ can again be neglected but the linear in $T$
part must be retained. Since ``hyperphotons'' are massless in the restored phase we
also
include the contribution of the corresponding ring diagrams:
\begin{equation}  \label{Y1ring} V_{restored}^{ring}(H_Y, T ) = - \frac{T}{12 \pi} [\frac{2}{3} (g'/2)^{2} T^2 +
m^2_{D_f} - \frac{((g'/2) H_Y)^{1/2} T}{2 \pi} - \frac{1}{8 \pi^2} (g'/2)H_Y ]^{3/2},
\end{equation}
where $m^2_{D_{f}}= \frac{1}{24} g'^2 T^2 \sum\limits_{f(R,L)}Y^2_f$ is the sum over 
the fermion
contributions to the Debye mass of the ``hyperphotons'', $Y_f$ are the hypercharges  of $R-$
and $L-$ leptons and quarks.
  Both these expressions have been calculated in a way described in the previous
sections.

For convenience of numerical investigations let us rewrite Eqs. (\ref{Yscal}) and  
(\ref{Y1ring}) in terms of the dimensionless variables $h,~ B $: $V(H_Y, T )_{restored}
= (H_0
)^2 v_{restored} (h, B )$,
\begin{eqnarray} \label{EPr} v_{restored} (h, B ) &=&\frac{1}{2} \frac{h^2}{
cos^2 \theta } +  \frac{\alpha}{3\sqrt{2} cos^3 \theta} \frac{h^{3/2}}{B}
\nonumber \\
&-& \frac{1}{3}\frac{\alpha}{B} [\frac{7}{6} \frac{4 \pi \alpha}{ cos^2 \theta B^2}  - \frac{h^{1/2}}{2\sqrt{2} \pi B cos \theta} - \frac{h}{16 \pi^2  cos^2 \theta} ]^{3/2},
\end{eqnarray}
where $\alpha = e^2/ 4 \pi$ and $h_Y = h/cos \theta$.

\section{Symmetry behaviour in a hypermagnetic field }

Now, let us investigate the EW phase transition in the hypermagnetic field for
 different
values of $m_H$. It can be done by considering the Gibbs free energy in the
broken, $G_{broken}(H^{ext}, \phi_c, T),$ and the restored,
$G_{restored}(H_Y^{ext}, T)$, phases \cite{Shap1}, \cite{Elm1}.
  The first order phase transition can be determined from two equations:
\begin{equation} \label{cond}
G_{restored}(H_Y^{ext}, T, 0) = G_{broken}( H^{ext}, T, \phi(H^{ext})_c ),
\end{equation}
describing the advantage of the broken phase  creation,
where $\phi(H)_c$ is a scalar field vacuum expectation value at given $H, T$
 which has
to be found as the minimum position of the total EP,
\begin{equation} \label{min}
\frac{\partial V( H, T, \phi_c)^{total}}{\partial \phi_c} = 0.
\end{equation}
Hence  the critical field strength can be calculated. In this expression (and below) we
write for brevity $H$ instead $H^{ext}$.

Having obtained the EP in the restored phase, the one-loop EP described by
formulae (\ref{12}), (\ref{14a}), (\ref{16})-(\ref{20}) and the ring diagram contributions
$V_{ring}$ we are
going to investigate the symmetry behaviour.
 First, we consider the total EP as the function of $\phi^2$ at
 various
fixed $H$, $T$, $K$ and determine the form of the EP curves in the broken
phase.
In this way it will be possible to select the range of the parameters when the
first order phase transition is realized. After that the
temperature  $T_c$ at given field strength $(H_Y)_c $ will be estimated
using Eqs. (\ref{cond}), (\ref{min}).

As usually \cite{Sk2}, to investigate symmetry behaviour we consider the
difference ${\cal V^{'}} = Re [{\cal V}(h,\phi, K, B) - {\cal V}(h, \phi= 0, B )]$
which  gives information about symmetry restoration.

In what follows it will be also convenient to express the conditions of the
phase transition in terms of the dimensionless variables $h, B, \phi,$ taking
 into account the relation $h_Y = h/cos \theta$. Then,  the Gibbs free energy
\begin{equation} \label{Gbrok}
G_{broken}(h^{ext}, \phi, B) = \frac{h^2}{2} + v^{'}(h, \phi, B) - h h^{ext},
 \end{equation}
has to be expressed through $h^{ext}$ by using the equation
\begin{equation} \label{cond1}
h^{ext} = h + \frac{\partial v^{'}(h, \phi, B)}{\partial h},
\end{equation}
where $v^{'}$ describes the one-loop and the ring diagram contributions to
 the EP.
The phase transiton happens when the relation holds,
\begin{equation} \label{T_c}
\frac{h^2}{2}tan^2\theta = v^{'}_{restored}(h, B_c) - v^{'}_{broken}(h, \phi_c, B_c).
\end{equation}
The function $v_{restored}^{'}$ is given by Eq. (\ref{EPr}). We also have
substituted the field $h^{ext}$ by $h$.

The results on the phase transition determined
by numerical investigation of the total EP are summarized in Table 1.
~ \\ ~
\begin{tabular}{c c c c c c c} \hline
~ \\ ~
~h&$K$&$T_c(GeV)$&$\phi_c(h,T_c)$&$\phi^2_c(h,T_c)$&$R$&$M_w(h,B_c)$ \\
[5pt] \hline
~0.01&0.85&106.47& 0.301662 & 0.091 & 0.69699 & 0.327235 \\
~0.01&1.25&122.21& 0.181659 & 0.033 & 0.36567 & 0.230086 \\   
~0.01& 2  &145.56& 0.094868 & 0.009 & 0.16033 & 0.186168 \\ \hline

~0.1 &0.85&108.58& 0.275681 & 0.076 & 0.62459 & 0.245186 \\
~0.1 &1.25&123.54& 0.130384 & 0.017 & 0.25963 & 0.112721 \\
~0.1 & 2  &148.39& 0.031623 & 0.001 & 0.05242 & 0.126315 \\ \hline

~0.5 &0.85&108.89& 0.248998 & 0.062 & 0.56253 & 0.49938 i\\
~0.5 & 1.25&second&order & phase &transition                         \\
~0.5 & 2   &second&order & phase &transition          \\ \hline
~ \\ ~
Table 1.
\end{tabular}

In the first column we show the hypermagnetic field strength in the
broken phase (in dimensionless units). In the second and third the
mass parameter $K = m^2_H/ M^2_w$ and the critical
temperature of the first order phase transition are adduced. Next two
columns give the local minimum positions $\phi_c(H, T_c)$  and their squard
values at the
transition temperatures. The last two columns fix the ratio $R =
246$ GeV $\phi_c(h, T_c)/ T_c $, determining the advantage of 
baryogenesis, and the $W$-boson effective mass calculated in the local
minimum of the EP at the corresponding field strengths and the transition
 temperatures.

As it is seen, the increase in $h$ makes the phase
transition weaker (not stronger as it was expected in  Refs. \cite{Shap1}, \cite{Elm1}
by analogy to superconductivity in the external magnetic
field). The ratio R is less than unit for  all the field strengths,
wherease the baryogenesis condition is $R > 1.2 - 1.5$ \cite{Shap2}.
Thus, we come to the conclusion that external hypermagnetic fields do
not make the EW phase
transition strong enough to produce baryogenesis.
Moreover, for strong fields the phase transition is of second order
 for all the values of $K$ considered. These are the main observations of
our numerical investigations.

 Let us continue the analysis of data in the Table 1. For the field
strengths $H > 0.1 - 0.5 H_0 (H_0 = M^2_w/e)$ the phase
transition is of  second or weak first-order. The W-boson effective mass
squared
(in dimensionless units) $M^2_w(\phi_c, h, B_c) = \phi^2_c
(h, B_c) - h + \Pi(h, B_c)$ is positive for $h = 0.01$ and $h = 0.1$.
Therefore, the local minimum is the stable state at the first order phase
transition. For stronger fields, when the second order phase transition
 happens, the effective $W$-boson mass becomes imaginary. This reflects
the known instability in the external magnetic field which exhibits
itself even when the radiation mass of the tachyonic mode is included.
 But it does not matter for the problem of searching for the strong first
order phase transition in the external hypermagnetic field investigated in the present
paper. The instability has to result in the condensation of $W$- and
$Z$-boson fields at high temperature.

In Refs. \cite{Shap1}, \cite{Shap4}, \cite{Elm1}, \cite{Lain}
it has been determined that the strong hypermagnetic field increases the
strength of the first order phase transition and in this case 
baryogenesis survives in the SM. Our results are in obvious
contradiction with this conclusion. To explane the origin
of the discrepancies let us first consider Refs. \cite{Shap1}, \cite{Elm1}
 where a perturbative method of computations has been applied. These
authors, considering the EW phase transition, have allowed for the
influence of the external field at tree level, only. That corresponds to the
usual case of superconductors in the external magnetic field, and, as a
consequence, they observed the strong first order phase transition.
 In fact, the type of the phase transition was just ussumed, since no
investigations of the EP curve with all the particles included for different $H_Y, T$ have been
carried out.
In the former paper, the qualitative estimate of the phenomenom considered was given,
wherease in the latter
one the quantitative analysis in one-loop approximation for the temperature
dependent part of the EP has been done. Actually,  in both these
investigations the influence of the external field was reduced to 
consideration of the
condition (\ref{T_c}) fixing the transition temperature.
 The role of fermions and $W$-bosons in the field
 was not investigated at all. However, as we have observed, the
fermions (heavy and light) are of paramount importance in the phase
transition dynamics. Just due to them the  EW phase transition becomes
of  second order in strong fields (for the values of $K$ when it is of
first order in weak fields).

In Refs. \cite{Shap4}, \cite{Lain} the phase transition was investigated
by the method combining the perturbation theory and  the lattice
simulations.  As the first step in this approach the static modes are
 maintained in the high temperature Lagrangian. The fermions are nonstatic
 modes and decoupled.
 So, no reflections of the fermion properties in the external fields and,
hence, no information on the
 EP curve could be derived in this way. The only fermion remainder
 is the t-quark mass entering the effective universal theory
\cite{Shap2}, \cite{Shap4}, \cite{Lain}. In our analysis, it has been
observed that not only heavy but also light fermions
are important in strong external fields at high temperature.
In fact, for various field strengths  the fermoins
with different masses are dominant and we
have taken account of all of them. Moreover, we have allowed for
all the  ring correlation corrections in the external field that also
influences symmetry behaviour.

We would like to notice that our  perturbative results for
the values of $K \sim 0.8 - 0.9$ are reliable. They are in agreement with
nonperturbative analysis at zero field. The external field is
taken into account exactly. For these masses of the Higgs particles we observed
the change of the first order phase transition to the second order one
 with increase in the field strength.
The same behaviour takes place for $K > 1$ when  parturbative analysis may
be not trusty. But, as we have discovered, the general picture of the field effects 
is only
quantitatively changed  for heavy scalar particles. In this case also the first
order phase transition in weak fields becomes of second order one for strong
 fields.  These circumstances convince us that the assumption of Ref.
 \cite{Shap1} that the hypermagnetic fields is able to make the weak first-order EW phase
transition strong enough is not aproved by the detailed calculations.

\section{Discussion}

 In  Refs. \cite{Shap1}, \cite{Elm1}
 the influence of strong external hypermagnetic field on the EW phase transition
has been taken into account in tree approximation. Further studying of the
 phenomenon, naturally, has to allow for the radiation and correlation
corrections. This is the problem that we have
addressed to in the present paper. The main idea was to
determine the form of the EP curve in the broken phase and find the
range of the parameters $H_Y, K $ when  the EW phase transition is of first
order.  To elaborate that the consistent EP including the
one-loop and ring diagrams of all the fundamental particles has been
constructed. As  we have seen, the role of fermions and ring diagrams
in the external field is crucial when the structure of the broken phase
is described. The external field was taken into consideration exactly through Green's
functions. The minimum of the EP was found to be stable at
sufficently high temperatures when the first order phase transition
happens. This important property is fulfilled when the ring diagrams of  the
tachyonic mode are included. As a result, no conditions for $W$- and
$Z$-boson condensates are realized at high temperatures at the first
order phase transition. The condensates could be generated for stronger
fields at the second order phase transition. But in this case baryogenesis does not
survive. 

The influence of strong magnetic fields on the vacuum at high temperature is a complicate
corporative effect described by the total EP. At some chosen values of $H, T, K$ the
different  terms of it are dominant. To better understand the role of
fermions in symmetry behaviour let us adduce two terms of the
asymptotic expansion of the EP in the limit of $T \rightarrow \infty, H \rightarrow
\infty$. The first one is the term
$\sim H^2 log \frac{T}{m_f}$. Due to this term the light fermions are
dominant at high temperature.
The second term can be derived from the expansion of the zero temperature part
Eq. (\ref{19}). This expression  side by side with the leading term $\sim H^2
log\frac{eH}{m_f}$, which due to a "dimension parameter trading" is replaced by the above
written term, contains the subleading one $\sim - eH m_f^2 log\frac{eH}{m_f^2}$ (for
details see Ref. \cite{Ditt2}). This term acts to make "heavier" the Higgs particles 
in the field.  As a result, the second order temperature phase transition  is stimulated
 due to strong fields. As it is well known at zero field, the correlation corrections relax
the strength of the first order phase transition \cite{Car} - \cite{Kol}. This
property is provided by the structure of the $V_{ring}$ term of the EP (\ref{21}), 
independently of the field presence. The field-dependent terms of the Debye 
masses of the scalar (\ref{23}), photon (\ref{31}) and $Z$ particles are negative 
that decreases the mass values. Therefore, the field acts to make weaker the effect
of correlations as the zero field case is compared. But  nevertheless the relaxation 
effect as such holds.
 We also have investigated the influence of different parts of the EP on
symmetry behaviour. It was discovered that the change of the phase
transition kind  with increase in $H$ is due to the fermion temperature part of the 
EP. These remarks help us to have a notion about the role of fermions and
correlations in strong fields.

In papers \cite{Shap4}, \cite{Lain} the EW phase transition in the
hypermagnetic field has been investigated by means of the method combining
perturbation theory and lattice simulations and the main
conclusions of Refs. \cite{Shap1}, \cite{Elm1} were supported. 
 In the former two papers, because of
peculiarities of the calculation procedure adopted, the effects of the external field
due to fermions as well as the correlation corrections have not been allowed for.
 So, from the point of view of the present analysis
these results also do not reproduce correctly the  behaviour of the EP
curve in the broken phase.

The values of the Higgs boson mass investigated in the present paper correspond to
the cases when perturbative results are reliable (K = 0.85) and may be
not trusty (K = 1.25 , 2). However, since the external field is taken
into account exactly its effects  do not depend on the specific  K
values. As we have seen, an increase in $H_Y$ makes the EW phase
transition of second order for the field strengths $H_Y \sim 0.5 \cdot
10^{24}G$ for all the mass values investigated. For weaker fields the
phase transition is of first order but the ratio $R = \phi_c(H,T_c)/T_c$
is less then unit.  Hence we conclude that baryogenesis does not survive in the
minimal Standard Model in the smooth external hypermagnetic fields.

The authors thank M. Bordag and A. Linde for interesting discussions and
remarks and A. Batrachenko for checking of the numeric calculations.  One of
us (V.S.) grateful DFG grant No 436 UKR 17/24/98  for financial support and
colleagues from Institute of Theoretical Physics at Leipzig university  for
 hospitality.


\begin{thebibliography}{90}
\bibitem{Enq}    K. Enqvist, Int. J. Mod. Phys. D7 (1998) 331;
                            astro-ph/9707300 28 July 1997
\bibitem{She}    M. Sher,  Phys. Rep. 179 (1989) 273.
\bibitem{Esp}    J.R. Espinosa, Preprint DESY 96-107 IEM-FT-96-133, June 1996.
\bibitem{Shap1}    M. Goivannini and M. Shaposhnikov, Phys. Rev. D57 (1998) 2186.
\bibitem{Elm1}       P. Elmfors, K. Enquist and K. Kainulainen, Phys. Lett. B440,   
(1998)) 269;  hep-ph/9806403.
\bibitem{Shap4}  K. Kajantie, M. Laine, J. Peisa, K. Rummokainen and M.
                 Shapohnikov, Nucl. Phys. B544 (1999) 357; hep-lat/ 9809004.
\bibitem{Lain} M. Laine, hep-ph/9902282.
\bibitem{Come} D. Comelli, D. Grasso, M. Pietroni and A. Riotto, The sphaleron 
          in a magnetic field and electroweak phase transition, hep-ph/9903227.
\bibitem{Car}    M. Carrington, Phys. Rev. D45 (1992) 2933.
\bibitem{Din}  M. Dine, R.G. Leigh, P. Huet, A. Linde and D. Linde, Phys. Rev. D46,
(1992) 550. 
\bibitem{Kol}   M. Gleiser and E.W. Kolb, FERMILAB-Pub-92/222-A, NSF-ITP-92-
               102, June 1992
\bibitem{Rez}    Yu.Yu. Reznikov and V.V. Skalozub, Sov. J. Nucl. Phys. 46
                (1987) 1085. 
\bibitem{Ulim_H} V. Skalozub and M. Bordag, Mod. Phys. Lett. A (be published);
                hep-ph/9904333.
\bibitem{Cabk}   A. Cabo, O.K. Kalashnikov  and A.E. Shabad, Nucl. Phys. B185
(1981) 473.
\bibitem{SVZ1}   A.O.Starinets, S.A. Vshivtsev and V.Ch. Zhukovskii,
                 Phys. Lett. B322 (1994) 287. 
\bibitem{SVZ2}   A.S. Vshivtsev, V,Ch. Zhukovsky and A.O. Starinets,
                 Z. Phys. C61  (1994) 285.
\bibitem{Kal}   O.K. Kalashnikov, Fortschr. Phys. 32 (1984) 325.
\bibitem{Buch}  W. Buchmuller and O. Philipsen, Phys. Lett. B397 (1997) 112. 
\bibitem{Shap2}  V.A. Rubakov and M. E. Shaposhnikov, Uspehi Fiz. Nauk, 166 (1996)
493; CERN-TH/96-13, hep-ph/9603208 v2  10 Apr 1996.
\bibitem{Cha}    J. Chakrabarti, Phys. Rev. D28 (1983) 2657; ibd D29 (1984) 1859.
\bibitem{Reu}    M. Reuter and W. Dittrich, Phys. Lett. 144B (1984) 99.
\bibitem{Fuj}    J. Fujimoto and T. Fukuyama, Z. Phys. C19 (1983) 11 .
\bibitem{Sk1}    V.V. Skalozub, Sov. J. Nucl. Phys. 45 (1987) 1058.
\bibitem{Sch}    J. Schwinger, Phys. Rev. 82 (1951) 664.
\bibitem{AO}     J. Ambj$\ddot o$rn and P. Olesen, Nucl. Phys. B315 (1989) 606;
                 B330 (1990) 193.
\bibitem{MDT}    S. MacDowell and O. T$\ddot o$rnquist, Phys. Rev. D45 (1992) 3833.
\bibitem{Tak}    K. Takahashi,  Z. Phys. C26 (1985) 601.
\bibitem{Cab}    A. Cabo, Fortschr. Phys. 29 (1981) 495.
\bibitem{VZM}    A.S. Vshivtsev, V. Ch. Zhukovskii and B. V. Magnitskii, 
                DAN USSR, 314 (1990) 175; A. V. Borisov, A. S. Vshivtsev, V.
Ch. Zhukovskii and P. A. Eminov, Uspehi Fiz. Nauk, 167 (1997) 241.
\bibitem{Elp}   P. Elmfors and D. Persson, Nucl. Phys. B 538 (1999) 309;
 hep-ph/9806335.
\bibitem{Ditt} W. Dittrich, Wu-y. Tsai and K.-H. Zimmermann, Phys. Rev. D19 (1979) 2929.
\bibitem{Ditt2} W. Dittrich and M. Reuter, $Effective~ Lagrangians~ in~ Quantum~
Electrodynamics$, Lecture Notes in Physics, v. 220, (Springer-Verlag, 1985).
\bibitem{Elm}    P. Elmfors, D. Persson and B-S. Skagerstam, Phys. Rev. Lett.
                 71 (1993) 480.
\bibitem{SS}   V.V. Skalozub and A. V. Strel'chenko, In Proceedings of IV Int. Conf.
                On Quantum Field Theory under external Conditions, 8 - 14 September, 1998,
               Leipzig, Germany (submitted for publication to Yadernaya Fizika).
\bibitem{Sk2}    V.V. Skalozub, Sov. J. Part. Nucl. 16 (1985) 445.
\bibitem{Sk78}  V. V. Skalozub, Sov. J. Nucl. Phys. 28 (1978) 113.
\bibitem{Niol}  N. K. Nielsen and P. Olesen, Nucl. Phys. B144 (1978) 376.
\bibitem{Sk3}    V.V. Skalozub, Int. J. Mod. Phys. A 11 (1996) 5643.

\end{thebibliography}
\end{document}